\documentclass[conference]{IEEEtran}
\IEEEoverridecommandlockouts
\usepackage{cite}
\usepackage{amsmath,amssymb,amsfonts}
\usepackage{graphicx}
\usepackage{textcomp}
\usepackage{xcolor}

\def\BibTeX{{\rm B\kern-.05em{\sc i\kern-.025em b}\kern-.08em
    T\kern-.1667em\lower.7ex\hbox{E}\kern-.125emX}}
\usepackage{url}
\usepackage{graphicx,subfigure}
\usepackage{epstopdf}
\usepackage{amsmath}
\usepackage{algpseudocode}
\usepackage{amsmath}
\usepackage{amssymb}
\usepackage{amsthm}
\usepackage{caption}
\usepackage{booktabs}
\usepackage{enumitem}
\usepackage{amsfonts}
\usepackage{multirow}
\usepackage[normalem]{ulem}
\useunder{\uline}{\ul}{}
\usepackage{color}
\usepackage{array}
\usepackage{acro}
\usepackage{listings}
\usepackage[underline=true]{pgf-umlsd}

\usepackage{setspace}

\DeclareRobustCommand{\acrodef}[2]{\DeclareAcronym{#1}{short=#1,long=#2}}
\acrodef{RRC}{Radio Resource Control}
\acrodef{ZMQ}{ZeroMQ}
\acrodef{LSTM}{Long Short-Term Memory}
\acrodef{LAL}{Listen-and-Learn}
\acrodef{TS}{Technical Specifications}
\acrodef{NR}{New Radio}
\acrodef{eMBB}{enhancing Mobile Broadband}
\acrodef{mMTC}{massive Machine Type Communications}
\acrodef{O-RAN}{Open Radio Access Network}
\acrodef{NAS}{Non-Access Stratum}
\acrodef{AS}{Access Stratum}
\acrodef{UE}{User Equipment}
\acrodef{MITM}{man-in-the-middle}
\acrodef{OTA}{over-the-air}
\acrodef{CN}{core network}
\acrodef{gNB}{gNodeB}
\acrodef{DoS}{Deny of Service}
\acrodef{BS}{Base Station}
\acrodef{NSA}{Non Standard-Alone}
\acrodef{SA}{Standard-Alone}
\acrodef{EPS}{Evolved Packet System}
\acrodef{eNB}{evolved NodeBs}
\acrodef{AKA}{authentication and key agreement}
\acrodef{EC-AKA}{Ensured confidentiality Authentication and Key agreement}
\acrodef{ASME}{Access Security Management Entity}
\acrodef{UP}{User Plane}
\acrodef{KDF}{key derivation function}
\acrodef{IMSI}{international mobile subscriber identity}
\acrodef{S-TMSI}{SAE Temporary Mobile Subscriber Identity}
\acrodef{TLS}{Transport Layer Security}
\acrodef{LTE}{Long Term Evolution}
\acrodef{MME}{Mobility Management Entity}
\acrodef{HSS}{Home Subscriber Server}

\begin{document}


\title{Formal and Fuzzing Amplification: Targeting Vulnerability Detection in 5G and Beyond}


\author{
	\IEEEauthorblockN{Jingda Yang\IEEEauthorrefmark{1}, {\it Student Member, IEEE} and 
	Ying Wang\IEEEauthorrefmark{2}, {\it Member, IEEE}}
	\IEEEauthorblockA{\IEEEauthorrefmark {0} School of Systems and Enterprises, Stevens Institute of Technology, Hoboken, USA \\}
	\IEEEauthorblockA{\IEEEauthorrefmark{1}jyang76@stevens.edu, and \IEEEauthorrefmark{2}ywang6@stevens.edu}
}

\maketitle

\begin{abstract}
Softwarization and virtualization in 5G and beyond require rigorous testing against vulnerabilities and unintended emergent behaviors for critical infrastructure and network security assurance. Formal methods operates efficiently in protocol-level abstract specification models, and fuzz testing offers comprehensive experimental evaluation of system implementations. In this paper, we propose a novel framework that leverages the respective advantages and coverage of both formal and fuzzing methods to efficiently detect vulnerabilities from protocol logic to implementation stacks hierarchically. The detected attack traces from the formal verification results in critical protocols guide the case generation of fuzz testing, and the feedbacks from fuzz testing further broaden the scope of the formal verification. We examine the proposed framework with the 5G Non Standard-Alone (NSA) security processes, focusing on the Radio Resource Control (RRC) connection process. We first identify protocol-level vulnerabilities of user credentials via formal methods. Following this, we implement bit-level fuzzing to evaluate potential impacts and risks of integrity-vulnerable identifier variation. Concurrently, we conduct command-level mutation-based fuzzing by fixing the assumption identifier to assess the potential impacts and risks of confidentiality-vulnerable identifiers. 
During this approach, we established $1$ attack model and detected $53$ vulnerabilities. The vulnerabilities identified used to fortify protocol-level assumptions could further refine search space for the following detection cycles. Compared to the state of art fuzz testing, this unified methodology significantly reduces computational complexity, transforming the computational cost from exponential to linear growth. Consequently, it addresses the prevalent scalability challenges in detecting vulnerabilities and unintended emergent behaviors in large-scale systems in 5G and beyond.
\end{abstract}


\section{Introduction}

 
 Verticals in 5G and next-generation infrastructure create a diverse and intricate environment consisting of software, hardware, configurations, instruments, data, users, and various stakeholders~\cite{Alcaraz-Calero2018Leading5G-PPP}. With system complexity and its lack of security emphasis by domain scientists, the formed ecosystem requires a comprehensive evaluation and in-depth validation for improving transitional Critical Infrastructure (CI) security posture~\cite{shatnawi2022digital}. 


Two major state-of-the-art approaches, formal verification and fuzz testing, have been proposed to detect various vulnerabilities and unintended emergent behaviors of the 5G network. Formal verification provides a high-level protocol concept and logical proof of security and vulnerability~\cite{peltonen2021comprehensive}. For example, Hussian\cite{Hussain20195Greasoner:Protocol} et al. proposed a cross-layer formal verification framework, which integrates model checkers and cryptographic protocol verifiers by applying the abstraction-refinement principle. In contrast, fuzz testing can offer a detailed and comprehensive experimental evaluation and detect potential vulnerabilities in the 5G code implementation platform~\cite{Klees2018EvaluatingTesting} and has been proven to be successful in discovering critical security bugs in implemented software\cite{bratus2008lzfuzz}. However, the limitations of conventional pick-and-choose fuzz testing and formal analysis still exist, especially with the large-scale software stacks in the system. Given that each approach possesses unique strengths, we have proposed a tandem connection between the fuzz testing and formal methods to achieve a more comprehensive vulnerability detection and enable high assurance in the security analysis.

Further, we propose an integrated framework fortified by a reinforcing loop to detect vulnerabilities and unintended emergent behaviors in system design and implementations. Our approach advocates for a harmonized application of fuzz testing and formal analysis, aiming to establish a symbiotic cycle between these two methods. This integrated strategy is designed to facilitate the identification of vulnerabilities throughout the entire search space, thereby providing a comprehensive and robust mechanism for vulnerability detection. Formal verification provides valuable guidance and assumptions in reducing and directing fuzz testing. Conversely, fuzz testing broaden formal verification's scope by classifying uncertainty areas. Importantly, this integrated approach enables mutual amplification between the two methodologies. Following this approach, we discover multiple vulnerabilities due to absence of rudimentary \ac{MITM} protection within the protocol, which is unexpected considering that the \ac{TLS} solution to this issue has been in existence for well over a decade~\cite{berbecaru2023tls}. Our framework, characterized by robust automation, scalability, and usability, promises to enhance security assurance and resilience across both infrastructure and domain levels, striving to guarantee the absence of additional security issues within the system.
Additionally, the proposed approach could be applied to various open programmable communication platforms\cite{O-RANAlliance2018O-RAN:RAN,SoftwareRadioSystems2021SrsRANSRS}
Our major contributions are summarized below:
\begin{itemize}[noitemsep,topsep=5pt]
    \item We designed and implemented the integrated formal guided fuzz testing framework, which significantly improves efficiency and achieves scalability for large-scale 5G systems and discovery of new and exploited vulnerabilities in the \ac{NSA} 5G communication authentication process.
    \item We performed in-depth formal analysis on the \ac{NSA} 5G authentication process by converting informal protocols into a symbolic flowchart (Fig.~\ref{fig:lte_arch}), enabling comprehensive formal analysis.
    \item We discover multiple vulnerabilities due to absence of rudimentary \ac{MITM} protection that needs to be addressed the 3GPP technical standards and protocols, despite the \ac{TLS} solution to this issue has been in existence for well over a decade.
    \item With the proposed integrated formal and fuzz testing framework, we connected vulnerabilities detected by formal analysis to real-life attack models and discovered new vulnerabilities.
\end{itemize}

The rest of the paper is organized as follows. Section \ref{system_overview} introduces the structure of our proposed framework. We then discuss our design and formal symbolic transfer of the \ac{NSA} 5G communication establishment process in Section \ref{formal_analysis}, followed by a detailed analysis and illustration of the formal verification results in Section \ref{attack_model}.
Then, we propose proven solutions for each detected formal attack model, along with some novel suggestions. In Section \ref{fuzz_test}, we use the assumptions as a guide to apply our proposed fuzz testing framework. 
Lastly, in Section \ref{comparsion}, we use  mathematical proof to analyze the efficiency of different fuzzing strategies across varied scopes of fuzz testing.

\section{System Design}\label{system_desin}
\subsection{Architecture Overview}\label{system_overview}
\begin{figure}[!t]
    \centering
    \includegraphics[width=0.45\textwidth]{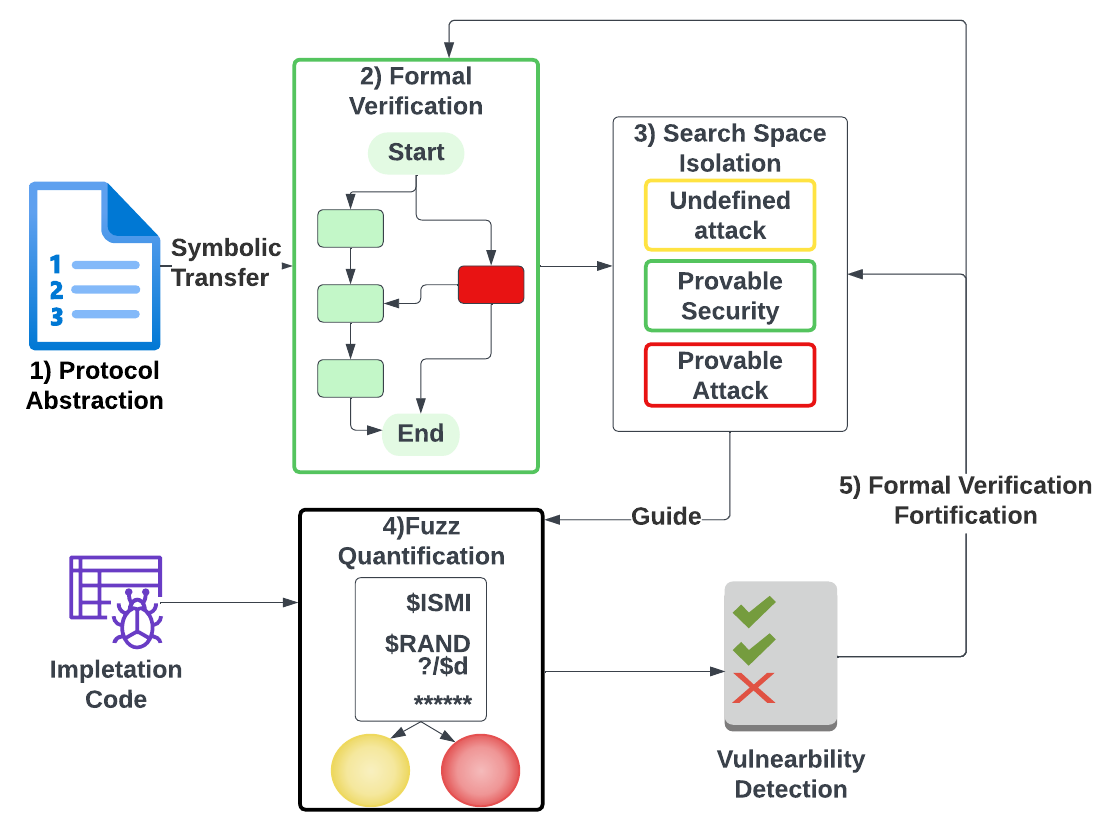}
    \vspace{-7pt}
    \caption{System Design}
    \label{fig:system_design}
    \vspace{-15pt}
\end{figure}

We design and implement a hybrid multi-model vulnerability and unintended emergent behaviors detection framework for 5G and other communication systems. As shown in Fig.~\ref{fig:system_design}, to achieve the amplification and cross-validation of fuzz testing and formal verification,  the proposed framework composites the following components to build up a reinforcing loop: 
    \begin{enumerate}[noitemsep,topsep=5pt]
        \item \textbf{Protocol Abstraction:} At the beginning of the system, we abstract the protocol into symbolic language. Logical transfer can easily exploit vulnerabilities in design. 
        \item \textbf{Formal Analysis:} In the formal verification process, we employed Proverif~\cite{SEC-004}, a robust tool, to conduct an in-depth analysis of our system's protocols. Proverif offers a logical proof of security properties and potential vulnerabilities, facilitating a robust and comprehensive evaluation of the system's security integrity. 
        \item \textbf{Search Space Isolation:} The output of formal verification divides the search space into three sets: no vulnerabilities, attack trace detected, and uncertain areas that need further investigation. The division of the search space effectively narrows down the uncertain regions and enables the scalability of vulnerability detection.
        \item \textbf{Formal Guided Fuzz Framework:} With the guidance of a formal verification conclusion, we initiate fuzz testing on runtime binary systems, focusing particularly on the predefined uncertain areas and those areas where attack traces have been detected. Fuzz testing serves a dual purpose: it is not only deployed to identify runtime vulnerabilities, thereby complementing the detection of vulnerabilities through logical proofs on protocols, but it also functions as a stochastic approach for those uncertain areas that cannot be verified through formal methods.
        \item \textbf{Fortification of Protocol and Formal Verification :} We verify the vulnerabilities detected by fuzz testing and feedback to the formal result and search space. By defining the space more precisely, formal verification can be further optimized, consequently extending the scope of the security assurance area.
    \end{enumerate}
    
   The proposed framework inter-connected with our previous fuzzing platform ~\cite{5grrrc}\cite{Wang2021DevelopmentResearchb} is capable of performing mutation-based identifiers fuzzing and permutation-based command fuzzing following the direction from the formal method conclusion. Formal verification, guided fuzzing analysis of results from the actual 5G testbed, and the real-time analysis and feedback construct a reinforcing loop in our system.

\subsection{Abstraction of \ac{NSA} 5G Authentication Protocol }\label{formal_analysis}
\begin{figure}[!t]
    \centering
    \includegraphics[width=0.5\textwidth]{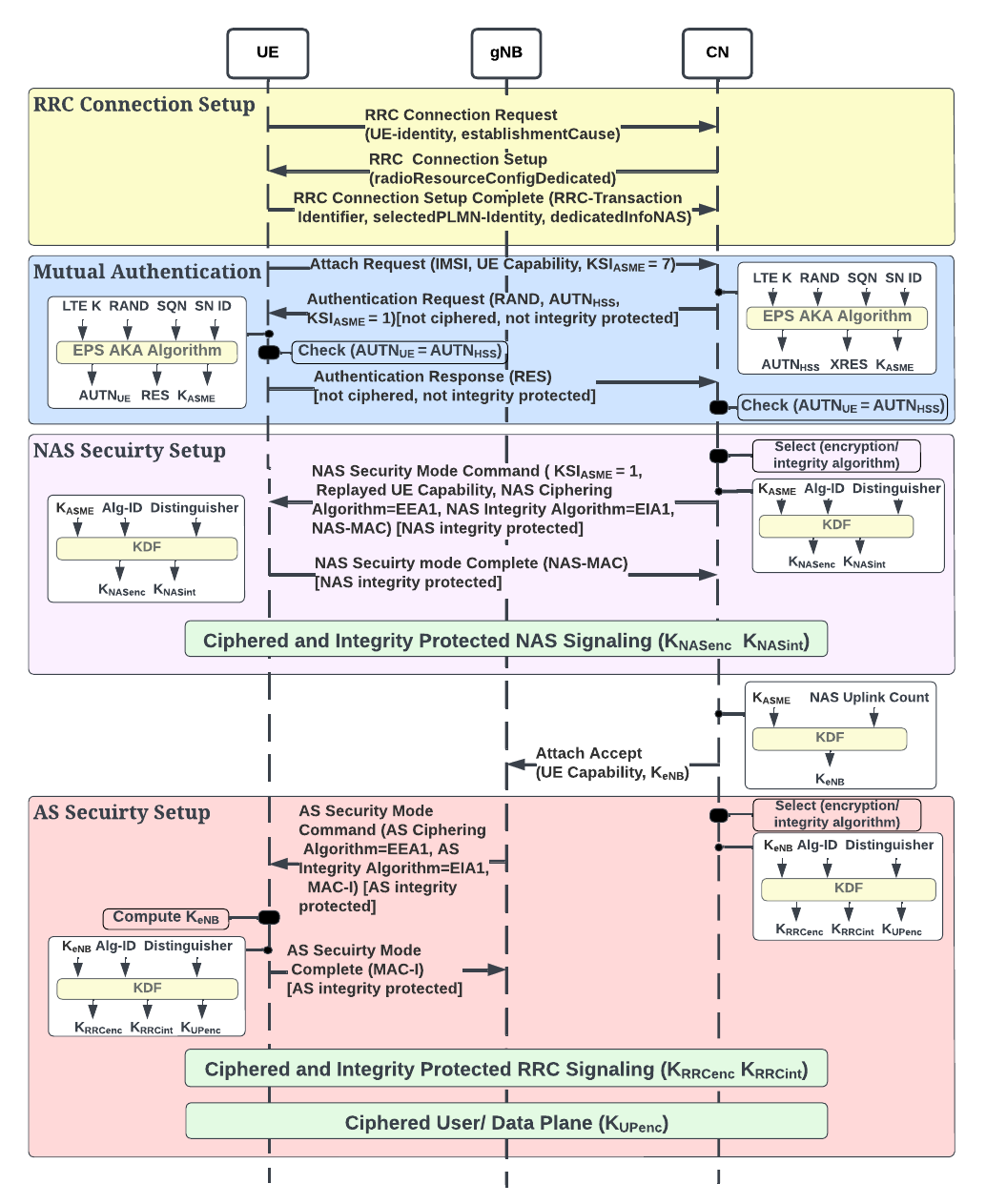}
    \caption{5G NSA Security Process.}
    \label{fig:lte_arch}
    \vspace{-15pt}
\end{figure}

Compared to \ac{SA} 5G network architecture, \ac{NSA} 5G architecture is still widely adopted but more vulnerable because the complexity introduced by the \ac{LTE} compatibility in protocol designs and infrastructure implementation, especially for authentication and authorization. Therefore, we focus on the authentication process in \ac{NSA} 5G architecture. As shown in Fig.~\ref{fig:lte_arch}, the abstracted protocol authentication process in \ac{NSA} architecture includes four parts: \textbf{\ac{RRC} Connection Setup}, \textbf{Mutual Authentication}, \textbf{\ac{NAS} Security Setup} and \textbf{\ac{AS} Security Setup}. Considering the scope of this paper and the critical level among them, we pilot on the \ac{RRC} connection setup for in-depth analysis.  

The \ac{RRC} Connection Setup is a pivotal step in the initial establishment of communication between a mobile device and the network in the \ac{LTE} and 5G \ac{NR} frameworks. 
This procedure is instigated by the network upon receiving a connection request from the \ac{UE}, commonly in response to an initiating event such as a call or data session initiation.
    \ac{RRC} connection setup process aims to build up connections in \ac{RRC} layer. 
    
Further, we abstract and derive the dependency table, presented as Table~\ref{tab:dependency_table_rrc}, from the defined protocol, considering four essential security properties: confidentiality, integrity, authentication, and accounting. Utilizing Table~\ref{tab:dependency_table_rrc}, we construct the corresponding dependency graph, as depicted in Fig~\ref{fig:dependency_graph_rrc}, to provide a visual representation of the security dependency relationships.

\subsection{Formal Guided Fuzz Framework}

Compared to traditional fuzz testing, which needs a complete understanding of code implementation, like LZFUZZ~\cite{bratus2008lzfuzz}, we propose a novel formal-guided identifier-based fuzzing framework. In our proposed fuzzing framework, we first fix the value of critical identifiers under the assumption proved by the formal verification and collect the communicated commands. Then we set up a relay attack mechanism on srsRAN platform~\cite{gomez2016srslte} following the attack traces, which are detected by formal verification. 


\section{Formal Detected Attack Model and Analysis }\label{attack_model}
In this section, we present a proof-of-concept via an illustrative attack model detected using Proverif~\cite{SEC-004}. A comprehensive summary of all identified attack models in 5G authentication and authorization process from our findings is depicted in Table~\ref{tab:summary}. We specifically focus on the \ac{RRC} connection setup for an in-depth demonstration.

\begin{table*}[]
\footnotesize
\begin{tabular}{|m{2.5cm}|m{2.5cm}|m{1cm}|m{1cm}|m{5cm}|m{3.5cm}|}
\hline
Attack &
  Vulnerability &
  Assumpt- ion &
  New attack?&
  Solution &
  Guidance to fuzz \\ \hline
  Modification of \ac{RRC} Connection &
  Modified commands will disable the \ac{RRC} functions &
  known C-RNTI or TMSI  &
  Inspired by~\cite{Hussain20195Greasoner:Protocol} &
  \begin{itemize}[leftmargin=*]
    \item Hash value protection for \ac{UE} identity
    \item Integrity protection
  \end{itemize}
  &
  Fuzz testing can start with different \ac{RRC} status. \\ \hline
  

 \ac{DoS} or Disconnect using Authentication Request. &
  \ac{UE} accepts authentication request without integrity. &
  No &
  Inspired by \cite{tsay2012vulnerability} &
        \begin{itemize}[leftmargin=*]
          \item \ac{EC-AKA}~\cite{6235098} 
          \item Hashed \ac{IMSI}~\cite{3gpp.29.118}
          \item Hashed \ac{IMSI} with integrity check~\cite{khan2018defeating}
      \end{itemize}
  &
  Repeat authentication request commands can be fuzzed at random time to test \ac{DoS} and cutting of device attack.\\ \hline

  Exposing $K_{NASenc}$ and $K_{NASint}$ &
  All \ac{NAS} information will be monitored, hijacked and modified. &
  known \ac{IMSI}, \ac{MITM} &
  Inspired by \cite{raza2018exposing} &
  \begin{itemize}[leftmargin=*]
    \item Asymmetric encryption
    \item Hashed \ac{IMSI} based encryption
  \end{itemize}&
  \ac{NAS} fuzz testing can start with known $K_{NASenc}$ and $K_{NASint}$. \\ \hline
  Exposing $K_{RRCenc}$, $K_{RRCint}$ and $K_{UPenc}$ &
  All \ac{RRC} and UP information will be monitored, hijacked and modified. &
  known \ac{IMSI}, MITM &
  Yes &
  \begin{itemize}[leftmargin=*]
    \item Asymmetric encryption
    \item Hashed \ac{IMSI} based encryption
  \end{itemize} &
  \ac{RRC} fuzz testing start with known $K_{RRCenc}$ and $K_{RRCint}$; User Plane (UP) fuzz testing start with known $K_{UPenc}$. \\ \hline
\end{tabular}
\caption{Summary of Findings}
\label{tab:summary}
\vspace{-15pt}
\end{table*}

\subsection{User Credentials Disclosure}\label{mutual_disclosure}
    In this attack, the adversary can exploit the transparency of \ac{RRC} Connection Setup process to effortlessly access critical user identity information, which includes but is not limited to the \ac{UE} identity and establishment cause. This illicit access enables the adversary to acquire user information and use the ensuing session key for nefarious activities such as eavesdropping and manipulation of subsequent communications.
    
\begin{figure}[!t]
        \centering
          \includegraphics[width=.45\textwidth]{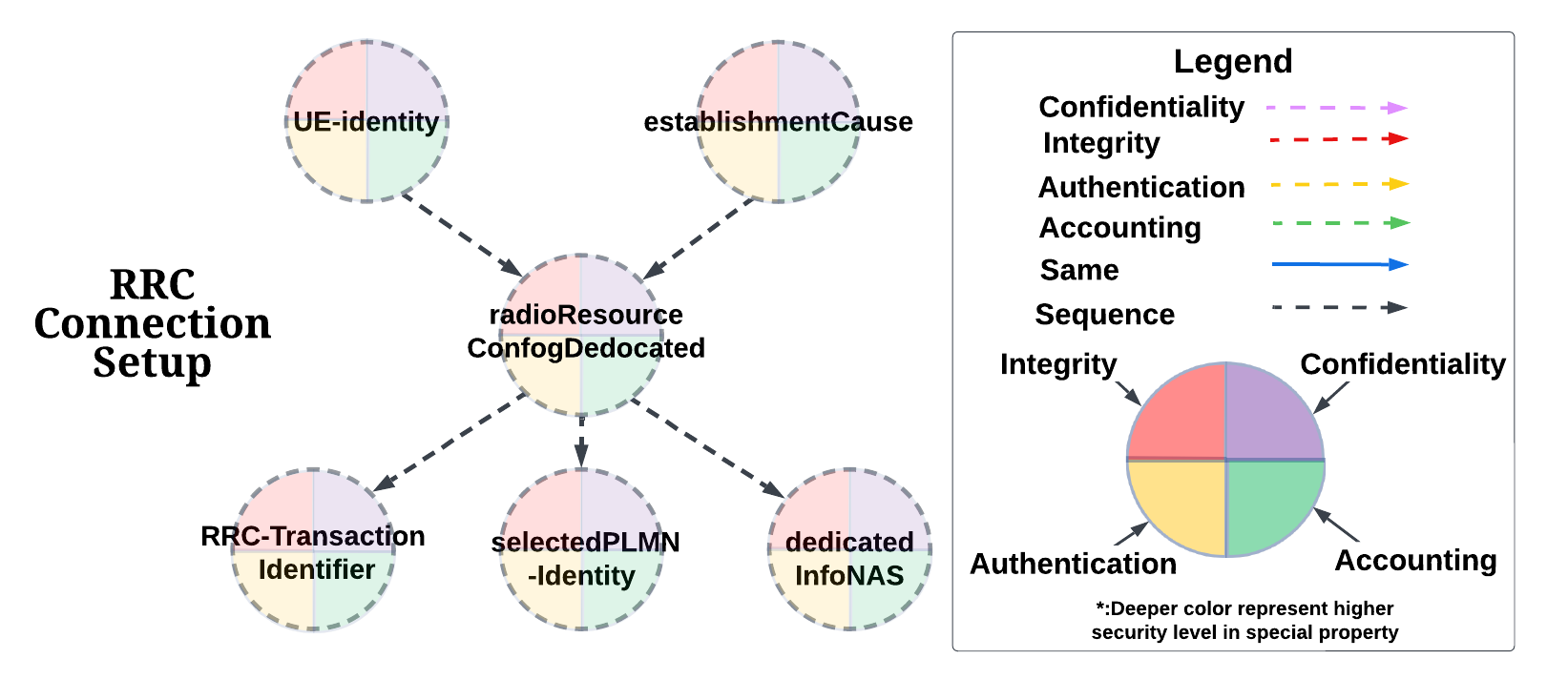}
          \caption{Dependency Graph of \ac{RRC} Connection Setup}
         \vspace{-15pt}
\label{fig:dependency_graph_rrc}
    \end{figure}

    \textbf{Assumption.} Analyzing Fig~\ref{fig:dependency_graph_rrc}, we can conclude that the adversary can exploit the transparency of \ac{RRC} Connection Setup process to directly access any identifier within the message. Furthermore, the adversary is also capable of establish a fake \ac{UE} or a \ac{MITM} relay to eavesdrop and manipulate the messages within the \ac{RRC} Connection Setup process. To verify the security properties of identifiers within the \ac{RRC} Connection Setup process, including aspects such as confidentiality and consistency, we converted the aforementioned assumptions into ProVerif code.
    
    \begin{table}
        \centering
          \includegraphics[trim={1cm 0 0 0},width=.42\textwidth]{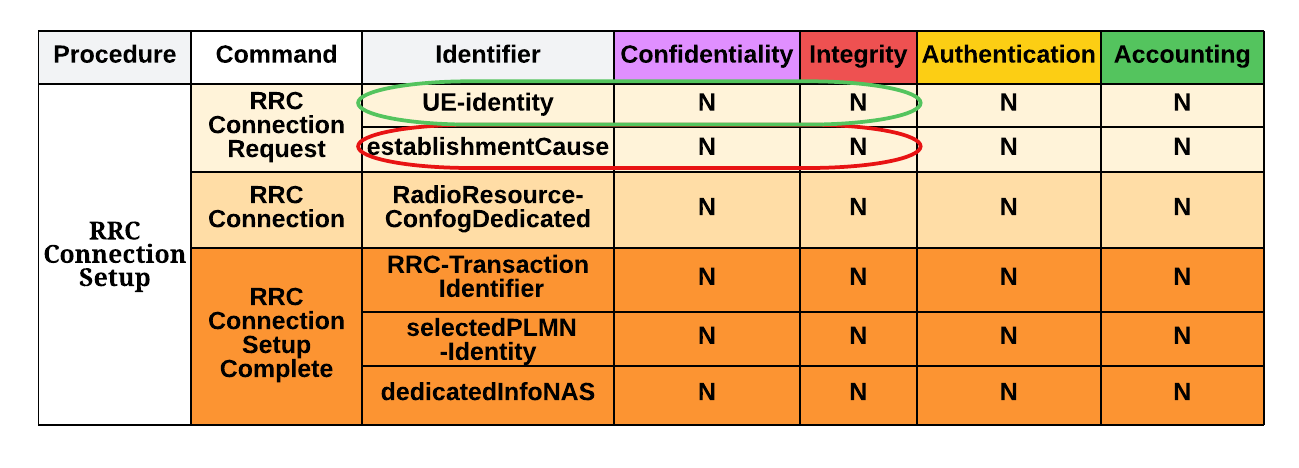}
          \vspace{-10pt}
          \caption{Dependency Table}
          \label{tab:dependency_table_rrc}
          \vspace{-8pt}
    \end{table}
    
    \textbf{Vulnerability.} As depicted in Fig.~\ref{fig:lte_arch}, the \ac{UE} initiates the process by sending an \ac{RRC} connection request to the \ac{CN}. Upon receiving this request, the \ac{CN} responds by transmitting the $radioResourceConfigDedicated$ back to the \ac{UE}. The \ac{UE}, in turn, obtains authentication from the \ac{CN} and responds with the $RRC-Transaction Identifier$, $selectedPLMN-Identity$ and $dedicatedInfoNAS$ to finalize the \ac{RRC} connection setup.
    Nevertheless, this process presents an exploitable vulnerability as an adversary can access all message identifiers. Such unprotected identifiers run the risk of being eavesdropped upon and modified, potentially enabling the adversary to orchestrate a \ac{MITM} relay attack.
    
\begin{figure}[!t]
        \centering
        \vspace{-8pt}
        \includegraphics[width=0.45\textwidth]{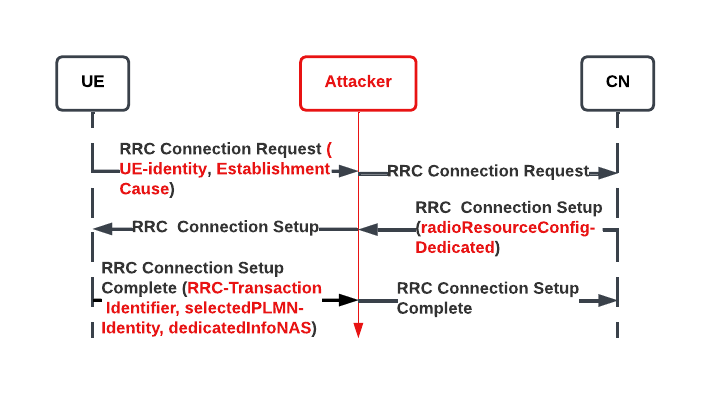}
        \vspace{-15pt}
        \caption{User Credentials Disclosure}
        \label{fig:user_identity}
        \vspace{-15pt}
    \end{figure}

    \textbf{Attack Trace Description.} Employing formal verification, we analyzed the confidentiality of identifiers within the \ac{RRC} Connection Setup process. Through this methodical investigation, we identified two categories of identifiers with the most significant impact: user identities and \ac{RRC} configuration identifiers. As illustrated in Fig.~\ref{fig:user_identity}, an attacker can access the identifiers marked in red, delineating the pathway of the attack.
    In the initial scenario, an adversary with the access to the user identity, like $UE-identity$, is capable of launch \ac{DoS} attack with real $UE-identity$. Contrary to traditional \ac{DoS} attacks, which aim to overwhelm a system's capacity, an $UE-identity$-based \ac{DoS} attack efficiently disrupts the \ac{CN} verification mechanism through repeated use of the same $UE-identity$, leading to authentication confusion.
    And in second case, with computationally derived $RRC-Transaction Identifier$, the adversary can establish a fake base station or perform a \ac{MITM} relay attack by manipulating these identifiers. In the latter case, the adversary positions between the \ac{UE} and the \ac{CN}, intercepting and modifying communications in real-time. Consequently, this attack model presents a severe threat to the security and integrity of the mobile network's communication.
        
    \textbf{Fortification via Formal Traced Vulnerability} 
    Given the significance and susceptibility of identifiers within the \ac{RRC} Connection Setup process, it is imperative to implement integrity protection measures for the $RRC-Transaction Identifier$. Additionally, adopting a hash value approach can assist in preventing the disclosure of \ac{UE} identity, further reinforcing security measures in this critical process.
    

\section{ Formal Guided Fuzzing Analysis}\label{fuzz_test}

As detailed in Section~\ref{attack_model}, formal verification delineates the system's security landscape into three zones: safe, non-safe, and undetermined. While the safe area necessitates no further scrutiny, the non-safe and undetermined areas warrant further investigation using fuzz testing. Specifically, we leverage fuzz testing to evaluate the impact of the non-safe areas within implementation stacks, as well as to ascertain the security level within the regions previously undetermined. By leveraging our previously proposed framework~\cite{yang2023systematic}, we effectively assess the security status of regions initially verified through formal methods. Due to the constraints of page length, we present a single example to illustrate the operation of our formally guided fuzzing framework. This example specifically demonstrates how the framework assess the impact of provable attacks that have been identified through formal verification.
    \subsection{\ac{MITM} bit-level fuzzing }
    In light of the identified vulnerabilities relating to confidentiality and integrity, we have developed a bit-level fuzzing test to examine the effects of exposed $UE-identity$ and $EstablishmentCause$. The results, as displayed in Table \ref{tab:rrc_request}, highlight two distinct outcomes. Modification of the $UE-identity$ has a minimal impact on authentication and communication, albeit with an introduction of some latency. Conversely, alterations to the $EstablishmentCause$ lead to a change in authentication types - a factor critical to the authentication establishment process, such as transforming an emergency request to data mode. There are total $8$ types of vulnerabilities that leverage the $EstablishmentCause$.
    Based on these bit-level fuzzing results, we can partition the provably insecure areas of the \ac{RRC} Connection Request into two categories: areas with less impact, including $UE-identity$, and areas with substantial impact, encompassing $EstablishmentCause$. Consequently, in subsequent fuzzing tests, we can strategically exclude $UE-identity$ fuzzing, focusing instead on the chain effects generated by high-risk identifiers.
        \begin{table}
        \centering
        
        \includegraphics[width=0.5\textwidth]{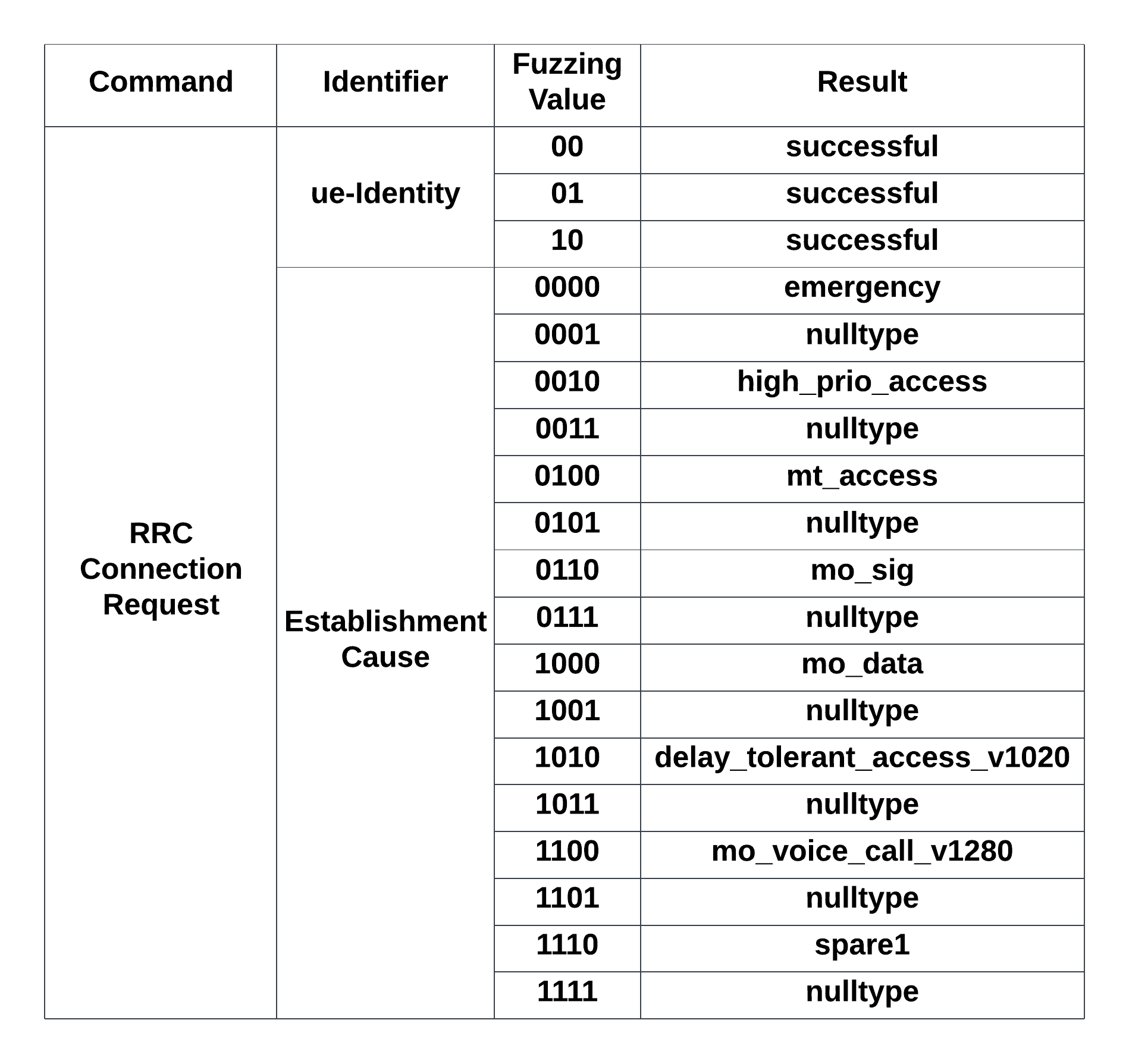}
        \vspace{-15pt}
        \caption{Fuzzing Result of User Credentials Disclosure}
        \label{tab:rrc_request}
        \vspace{-15pt}
        \end{table}
    \subsection{Command-level fuzzing}\label{rrc_commend}
    Assuming complete disclosure of all necessary \ac{UE} identities and an unprotected RAND in the Authentication Request of the Mutual Authentication process, it is a reasonable deduction that an adversary can acquire the RNTI, which is derived from \ac{UE} identities, and RAND, a crucial identifier for generating a session key. Unlike the boundless scenarios possible with black-box fuzzing, our approach uses a fixed session key to concentrate on the impact of a \ac{MITM} attack, thereby eliminating the computational waste associated with guessing random identifiers and \ac{UE} identities.

\begin{figure}[!htb]
        \centering
        \includegraphics[width=0.4\textwidth]{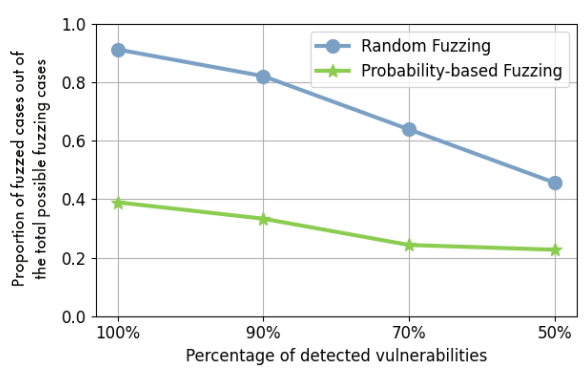}
        \vspace{-6pt}
        \caption{Comparison of Different Command-level Fuzzing Strategy Efficiency }
        \label{fig:Comparison_command}
        \vspace{-15pt}
    \end{figure}

    Building on our previously proposed probability-based fuzzing strategy \cite{yang2023systematic}, we have established a more efficient method for identifying unintended vulnerabilities that prove challenging to detect via formal verification. A comparison between random fuzzing and our probability-based approach (Fig. \ref{fig:Comparison_command}) reveals that our proposed probability-based framework requires only 36.5\% of the number of fuzzing cases used in a random fuzzing strategy to detect all 43 vulnerabilities~\cite{yang2023systematic}. 

\section{Performance and Efficiency Assessment}\label{comparsion}
Our proposed fuzz testing framework, guided by formal verification, affirms the viability of our integrative approach combining both formal and fuzz testing frameworks. In this section, we analyze the efficiency of fuzz testing and explore the relationship between formal verification and fuzz testing, underscoring the potential benefits of our innovative strategy. 

Fuzz testing is a methodical, brute-force approach to detecting vulnerabilities, accomplished by supplying an extensive range of random data to uncover potential security threats. However, due to computational constraints, exhaustive vulnerability detection for the entire 5G \ac{NSA} protocol, even for a singular command, is not practical. To increase the efficiency of fuzz testing, the rule-based mutation fuzz testing strategy has been proposed~\cite{Salazar20215Greplay:Injection}. This strategy refines the scope of fuzz testing to specific identifiers in line with protocol rules.

\begin{figure}[!htb]
        \centering\includegraphics[width=0.5\textwidth]{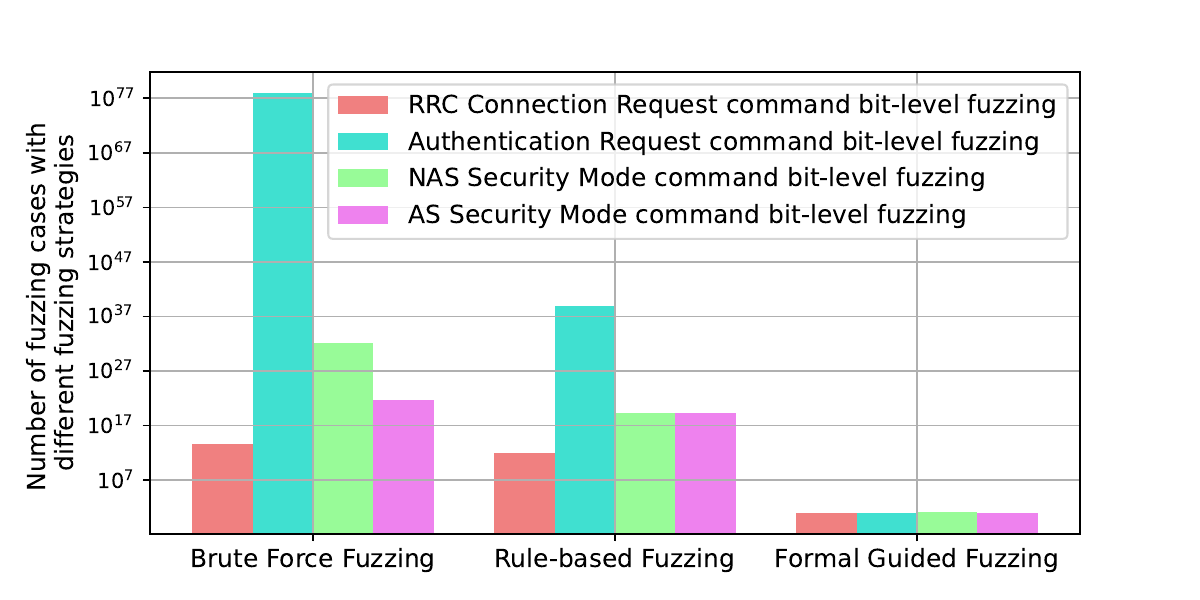}
        \vspace{-7mm}
        \caption{Comparison of Different Bit-level Fuzzing Strategy Efficiency }
        \label{fig:Comparison}
        \vspace{-4mm}
    \end{figure}

Although the rule-based mutation fuzz testing strategy yields a substantial reduction in computational complexity, it can still produce meaningless, randomly generated inputs. As a response, we introduce a formal-guided fuzz testing strategy. This strategy complies with formal verification assumptions and generates three categories of representative inputs: formal-based legal inputs, formal-based illegal inputs, and randomly generated inputs. While formal-based inputs must adhere to the protocol-defined rules or format, randomly generated inputs are not bound by these restrictions. The comparative efficiency of different fuzz strategies across four distinct processes is depicted in Figure \ref{fig:Comparison}. A detailed performance analysis of these varied fuzzing strategies is provided in the following section.

Based on the guidance of formal verification in Section~\ref{mutual_disclosure}, the \ac{RRC} Connection Request command, which includes 40 bits of UE-Identity, 4 bits of $EstablishmentCause$, and 1 bit of spare, is vulnerable to \ac{DoS} or \ac{MITM} attacks. Traditional brute-force fuzz testing generates more than $2^{45}$ fuzzing cases, and rule-based fuzzing generates $2^{40}+2^{4}+1$ fuzzing cases based on the defined identifiers. However, our formal guided fuzzing strategy requires only 9 fuzzing cases, including one legal UE-Identity case, one illegal UE-Identity case, one random out-of-rule UE-Identity case, $2$ legal/illegal $EstablishmentCause$ cases, one random out-of-rule $EstablishmentCause$ case, one legal spare case, one illegal spare case, and one out-of-rule spare case.

Our proposed framework has the capacity not only to validate the impact and security of identifiers, but also to detect unintended vulnerabilities based on high-risk assumptions, such as an identifier set that is accessible to an adversary. As corroborated by the evidence presented in Section \ref{rrc_commend}, our framework proves highly efficient in detecting vulnerabilities, underscoring its potential utility in enhancing system security.

\section{Conclusion}
In this paper, we have introduced an innovative framework that integrates formal verification and fuzz testing to fortify the security of 5G systems, effectively addressing the vulnerabilities from protocol logic to implementation stacks. The dynamic feedback loop within this framework has demonstrated its strength in both the refinement of undefined areas and the exhaustive detection of potential vulnerabilities. This work has been illuminated through an application on a continuous loop in the \ac{RRC} Connection Setup process, illustrating the practicability and effectiveness of our proposed methodology.
In the initial phase, our framework identifies a formal attack model through the application of formal verification. Subsequently, leveraging the protocol-level exposure of user credentials, the proposed framework employs bit-level and command-level fuzzing to execute comprehensive impact identification and simulate plausible attacks. As a result, by relying on the verified impact and the security status of the identifier or command determined by the fuzz test, our framework robustly reinforces protocol-level assumptions and refines the detection area. Notably, this integrated approach significantly mitigates computational complexity, transitioning it from exponential to linear growth. This scalability ensures that the framework can accommodate larger datasets or more complex scenarios without a drastic increase in computational resources or processing time, making it suitable for extensive applications in 5G security testing.

To conclude, our research presents a pioneering step towards bolstering 5G security by employing an integrated, hierarchical approach to vulnerability detection. This work contributes substantially to the ongoing efforts to secure the next generation of wireless communications and provides a foundation for future research in this domain. Further studies might explore extending this approach to other advanced wireless technologies to ensure robust security in our increasingly connected world.

\section*{Acknowledgment}
This effort was sponsored by the Defense Advanced Research Project Agency (DARPA) under grant no. D22AP00144. The views and conclusions contained herein are those of the authors and should not be interpreted as necessarily representing the official policies or endorsements, either expressed or implied, of DARPA or the U.S. Government.

\bibliographystyle{IEEEtran}
\bibliography{reference,references_ying}

\end{document}